\documentclass[aps,prl,twocolumn,superscriptaddress]{revtex4-1}
\usepackage{graphicx}
\usepackage[version=4]{mhchem}
\usepackage{siunitx}
\usepackage{color}
\usepackage{tabularx}
\usepackage{multirow}

\newcommand*{\fig}{Fig.\,}
\newcommand*{\eq}[1]{Eq.\,(#1)}
\newcommand*{\tab}{Table\,}
\newcommand*{\reftaken}{Ref.\,}
 
\newcommand*{\etal}{\emph{et~al. }}
\newcommand*{\ie}{\emph{i.e.\,}}
\newcommand*{\eg}{\emph{e.g.\,}}

\renewcommand{\vec}[1]{\boldsymbol{#1}}
\usepackage{changes}
\begin{document}

\title{Orientation Dependence of the Magnetic Phase Diagram of \ce{Yb2Ti2O7}}

\author{S. S{\"a}ubert}
\email[]{steffen.saeubert@colostate.edu}
\altaffiliation{Present address: Department of Physics, Colorado State University, Fort Collins, Colorado 80523-1875, USA.}
\affiliation{Physik-Department, Technische Universit{\"a}t M{\"u}nchen, D-85748 Garching, Germany}
\affiliation{Heinz Maier-Leibnitz Zentrum, Technische Universit{\"a}t M{\"u}nchen, D-85748 Garching, Germany}

\author{A. Scheie}
\affiliation{Institute for Quantum Matter and Department of Physics and Astronomy, Johns Hopkins University, Baltimore, Maryland 21218, USA}

\author{C. Duvinage}
\affiliation{Physik-Department, Technische Universit{\"a}t M{\"u}nchen, D-85748 Garching, Germany}

\author{J. Kindervater}
\affiliation{Institute for Quantum Matter and Department of Physics and Astronomy, Johns Hopkins University, Baltimore, Maryland 21218, USA}

\author{S. Zhang}
\affiliation{Institute for Quantum Matter and Department of Physics and Astronomy, Johns Hopkins University, Baltimore, Maryland 21218, USA}
\affiliation{Department of Physics and Astronomy, University of California, Los Angeles, California 90095, USA}

\author{H.J. Changlani}
\affiliation{Department of Physics, Florida State University, Tallahassee, Florida 32306, USA}
\affiliation{National High Magnetic Field Laboratory, Tallahassee, Florida 32304, USA}
\affiliation{Institute for Quantum Matter and Department of Physics and Astronomy, Johns Hopkins University, Baltimore, Maryland 21218, USA}

\author{Guangyong Xu}
\affiliation{NIST Center for Neutron Research, National Institute of Standards and Technology, Gaithersburg, Maryland 20899, USA}

\author{S.M. Koohpayeh}
\affiliation{Institute for Quantum Matter and Department of Physics and Astronomy, Johns Hopkins University, Baltimore, Maryland 21218, USA}
\affiliation{Department of Materials Science and Engineering, Johns Hopkins University, Baltimore, Maryland 21218, USA}

\author{O. Tchernyshyov}
\affiliation{Institute for Quantum Matter and Department of Physics and Astronomy, Johns Hopkins University, Baltimore, Maryland 21218, USA}

\author{C.L. Broholm}
\affiliation{Institute for Quantum Matter and Department of Physics and Astronomy, Johns Hopkins University, Baltimore, Maryland 21218, USA}
\affiliation{NIST Center for Neutron Research, National Institute of Standards and Technology, Gaithersburg, Maryland 20899, USA}
\affiliation{Department of Materials Science and Engineering, Johns Hopkins University, Baltimore, Maryland 21218, USA}

\author{C. Pfleiderer}
\affiliation{Physik-Department, Technische Universit{\"a}t M{\"u}nchen, D-85748 Garching, Germany}

\date{\today}

\begin{abstract}
In the quest to realize a quantum spin liquid (QSL), magnetic long-range order is hardly welcome. Yet it can offer deep insights into a complex world of strong correlations and fluctuations. Much hope was placed in the cubic pyrochlore \ce{Yb2Ti2O7} as a putative U(1) QSL but a new class of ultra-pure single crystals make it abundantly clear the stoichiometric compound is a ferromagnet. Here we present a detailed experimental and theoretical study of the corresponding field-temperature phase diagram. We find it to be richly anisotropic with a critical endpoint for $\vec{B}\,\parallel\,\langle 100\rangle$, while field parallel to  $\langle 110 \rangle$ and $\langle 111 \rangle$ enhances the critical temperature by up to a factor of two and shifts the onset of the field-polarized state to finite fields. Landau theory shows that \ce{Yb2Ti2O7} in some ways is remarkably similar to pure iron. However, it also pinpoints anomalies that cannot be accounted for at the classical mean-field level including a dramatic enhancement of $T_{\mathrm{C}}$ and reentrant phase boundary by fields with a component transverse to the easy axes, as well as the anisotropy of the upper critical field in the quantum limit.
\end{abstract}

\clearpage


\maketitle

\section*{Introduction}

Frustrating magnetism by affixing spins to lattices that are inconsistent with conventional magnetic order is a well-established route towards novel collective properties \cite{1994_Ramirez_AnnuRevMaterSci,2010_Gardner_RevModPhys}. When the interactions support quantum fluctuations, one may hope to indefinitely suppress magnetic phase transitions, replacing conventional forms of order and symmetry breaking with a quantum spin liquid and its emergent fractionalized quasi-particles \cite{2017_Savary_RepProgPhys,2019_Broholm_ArXiv}. Many frustrated magnets however, show fragile forms of magnetic order at low temperatures as well as extreme sensitivity to sample purity. To realize and document a QSL and to learn from proximate ordered phases, requires ultra-pure single crystalline samples and an array of comprehensive high-quality measurements in close coordination with theory. 

Here we report such a study of the quantum magnetism of \ce{Yb2Ti2O7}, a prototypical pyrochlore magnet \cite{2011_Yaouanc_PhysRevB,2011_Ross_PhysRevB,2012_Ross_PhysRevB,2017_Arpino_PhysRevB,2017_Mostaed_PhysRevB,2018_Ghosh_PhysRevB,2018_Sala_PhysRevB,2018_Shafieizadeh_SciRep,2019_Bowman_NatCommun}, in which we find despite geometric frustration and quantum fluctuations anisotropic ferromagnetism at low temperatures that appears to be deceptively simple at first sight. Early studies of \ce{Yb2Ti2O7} included a diffuse zero-field neutron spectrum \cite{2009_Ross_PhysRevLett,2017_Thompson_PhysRevLett,2016_Gaudet_PhysRevB,2018_Buhariwalla_PhysRevB}, as well as unconventional quasiparticles in the paramagnetic phase \cite{2019_Hirschberger_ArXiv,2016_Pan_NatPhys,2016_Tokiwa_NatCommun} which may be preserved to low temperatures in oxygen-deficient samples \cite{2019_Bowman_NatCommun}. More recently, an unusual reentrant field-dependent phase diagram was reported \cite{2017_Scheie_PhysRevLett} which has not yet been understood. 

For many years, it was thought that these unusual features of \ce{Yb2Ti2O7} signalled a quantum spin liquid (QSL) \cite{2011_Ross_PhysRevX,2013_Hayre_PhysRevB,2012_Applegate_PhysRevLett,2016_Pan_NatPhys,2016_Tokiwa_NatCommun,2017_Kermarrec_NatCommun} with long range entanglement and fractionalized excitations \cite{2004_Hermele_PhysRevB,2012_Savary_PhysRevLett,2012_Lee_PhysRevB,2014_Gingras_RepProgPhys,2019_Chern_SciRep}. However, in recent years the QSL hypothesis has lost favor because of the evidence of ferromagnetic order in  \ce{Yb2Ti2O7} \cite{2017_Scheie_PhysRevLett,2019_Bowman_NatCommun}, putative evidence for a structural instability \cite{2018_Trump_NatCommun}, and refined Hamiltonians that are inconsistent with a QSL \cite{2017_Thompson_PhysRevLett,2015_Robert_PhysRevB}. Instead, it has been proposed that the unusual features of \ce{Yb2Ti2O7} arise from a competition between ferromagnetism and antiferromagnetism \cite{2017_Yan_PhysRevB,2015_Jaubert_PhysRevLett}. Short-range correlations and exotic excitations above the magnetic ordering temperature indicate that this phase competition produces nontrivial effects, including a possible intermediate temperature QSL phase \cite{2019_Bowman_NatCommun,2015_Kato_PhysRevLett,2019_Castelnovo_PhysRevB}. 
Perhaps most intriguing, small angle neutron scattering and conventional neutron spectroscopy recently revealed evidence for a peculiar combination of splayed ferromagnetism with antiferromagnetic meso-scale textures as well as ferro- and antiferromagnetic spin waves \cite{2019_Scheie_ArXiv}. While this appears to suggest a near degeneracy of ferro- and antiferromagnetism, it raises as a key question, if and to what degree at least some component of these correlations may be captured with conventional concepts.

Here we focus on the uniform, static magnetization  complemented by susceptibility, specific heat measurements, and magnetic neutron diffraction. We explore the anisotropic field-temperature phase diagram of \ce{Yb2Ti2O7} as compared with the predictions of various standard models. We have examined the phase diagram for fields along each of the three main symmetry directions, $\langle 111 \rangle$, $\langle 110 \rangle$, and $\langle 100 \rangle$. For fields along $\langle 111 \rangle$, previously reported in \cite{2017_Scheie_PhysRevLett}, and $\langle 110 \rangle$ we find reentrant behavior, wherein an applied field initially increases the ordering temperature. For fields along $\langle 100 \rangle$, the high-field phase boundary collapses and the system enters a field polarized state for vanishingly small applied fields. All field directions show extremely small coercive fields, indicating essentially freely moving domain walls consistent with the high sample purity.

We compare our data with the predictions of a coarse-grained theoretical model that accounts qualitatively for our observations including the field-dependent magnetic structure that we infer from magnetic neutron diffraction. However, we also show significant discrepancies with classical mean-field calculations in the form of the orientation and temperature dependence of the upper critical field and of the large field-driven enhancement of the critical temperature (reentrance) for fields along $\langle 111 \rangle$ and $\langle 110 \rangle$. We speculate that these features are caused by the quantum fluctuations and/or collective physics of the underlying frustrated magnet beyond the mean-field approach. 

\section*{Results}

\subsection*{Experiments}

\begin{figure}[htbp]
	\includegraphics[width=0.792\columnwidth]{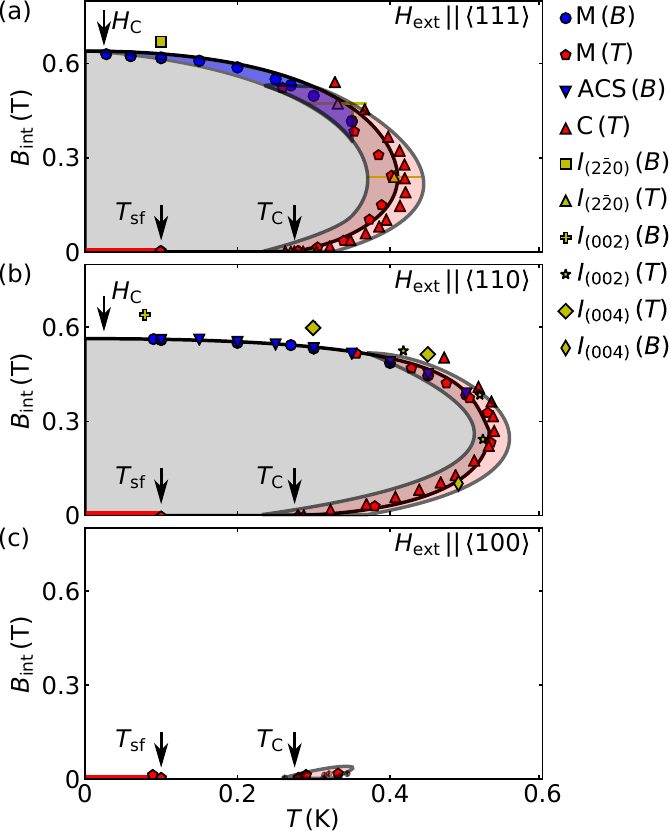}
	\caption{Magnetic phase diagram of \ce{Yb2Ti2O7} for applied fields along (a) $\langle 111 \rangle$, (b) $\langle 110 \rangle$, and (c) $\langle 100 \rangle$ as inferred from field and temperature dependent magnetization, specific heat, AC susceptibility, and neutron scattering. $(B)$ and $(T)$ indicate field and temperature scans, respectively. Hysteretic effects observed under field and temperature sweeps are indicated by means of blue and red shading, respectively.} 
	\label{fig:figure_1_phasediagram}
\end{figure}

\begin{figure}[htbp]
	\includegraphics[width=0.95\columnwidth]{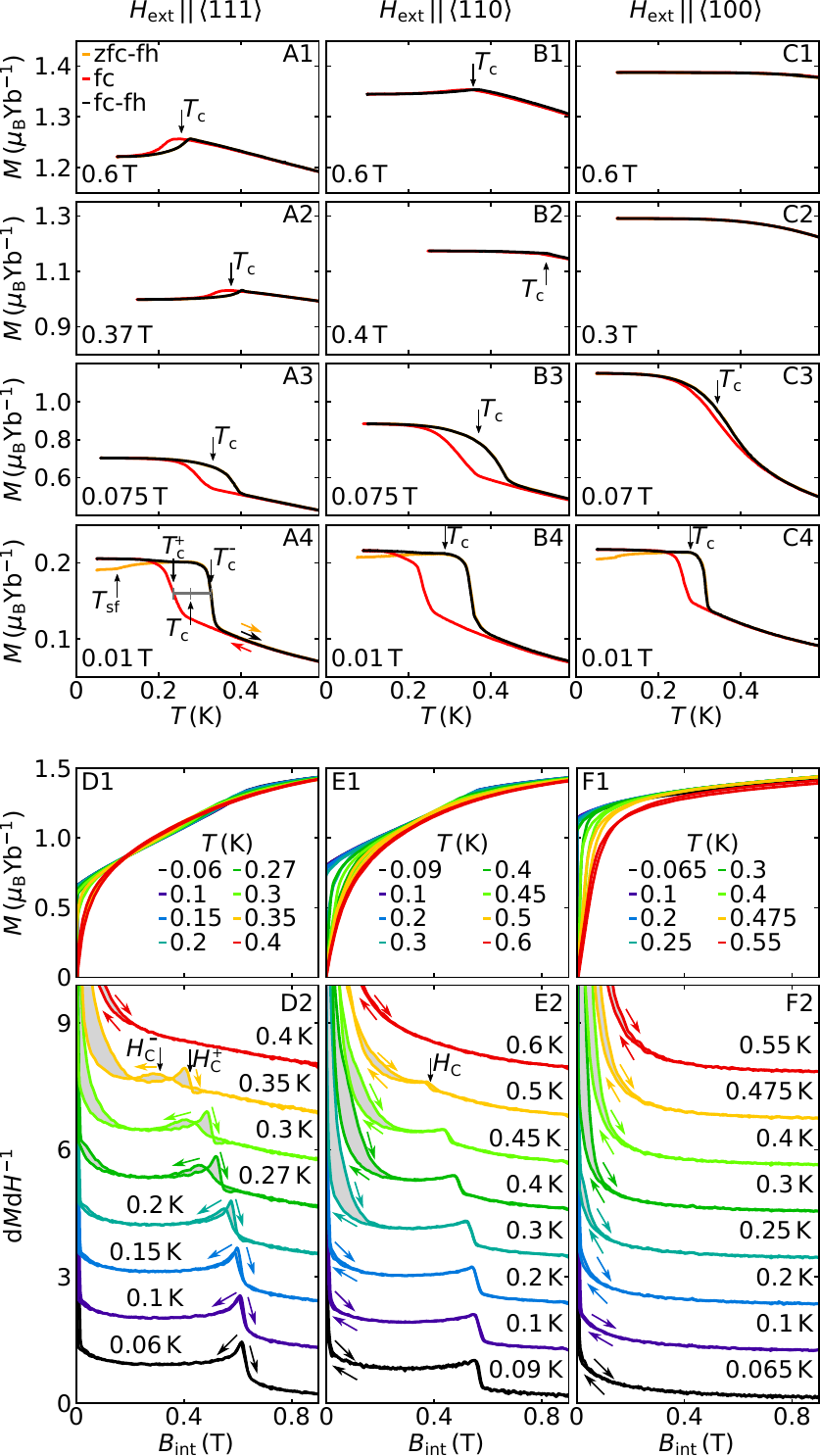}
	\caption{(A-C) Temperature dependence of the magnetization of \ce{Yb2Ti2O7} in high (A1-C1), intermediate (A2-C2 and A3-C3), and small (A3-C3) applied fields. In small applied fields and below \SI{100}{mK} a distinct difference between data recorded under zero-field-cooling (zfc) and field-cooling (fc) emerges, which has been attributed to to spin freezing in related rare-earth pyrochlore systems \cite{2012_Krey_PhysRevLett, 2012_Legl_PhysRevLett}.
	This feature vanishes for finite internal fields.
	(D-F) Magnetization and differential susceptibility of \ce{Yb2Ti2O7} as function of internal magnetic field after correction of demagnetization fields for the $\langle 111 \rangle$ (D), $\langle 110 \rangle$ (E), and $\langle 100 \rangle$ (F) direction, respectively. The differential susceptibility data are shifted with respect to each other for clarity.}
	\label{fig:figure_2_MvsT_MvsB}
\end{figure}

\begin{figure}[htbp]
	\includegraphics[width=\columnwidth]{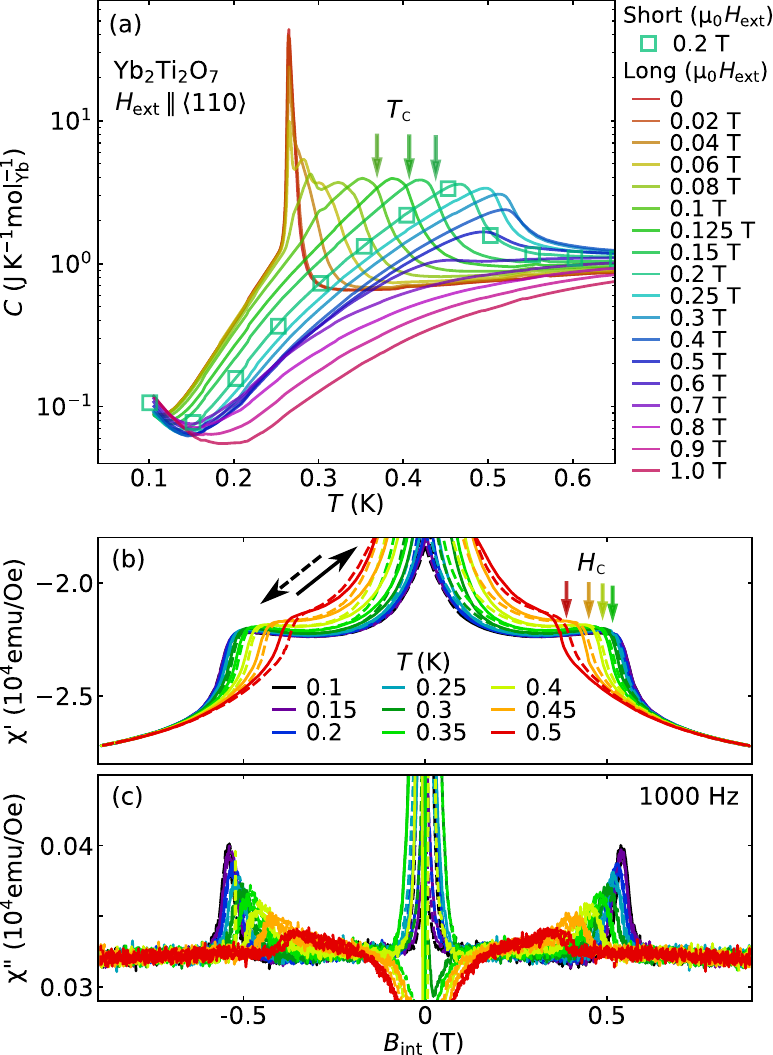}
	\caption{(a) Specific heat of \ce{Yb2Ti2O7} as a function of temperature at different $\langle 110 \rangle$ oriented magnetic fields. Note the sharp first-order-like anomaly at zero field which broadens and becomes a second-order lambda-like anomaly at finite field.
	(b-c) Real and imaginary components of the AC susceptibility as a function of $\langle 110 \rangle$ field at different temperatures. The negative field sweeps show invariance of field sweep direction for non-zero internal fields. Solid lines indicate increasing magnetic field and dashed lines show decreasing magnetic field.}
	\label{fig:figure_3_HC_ACDR}
\end{figure}

\begin{figure}[htbp]
	\includegraphics[width=\columnwidth]{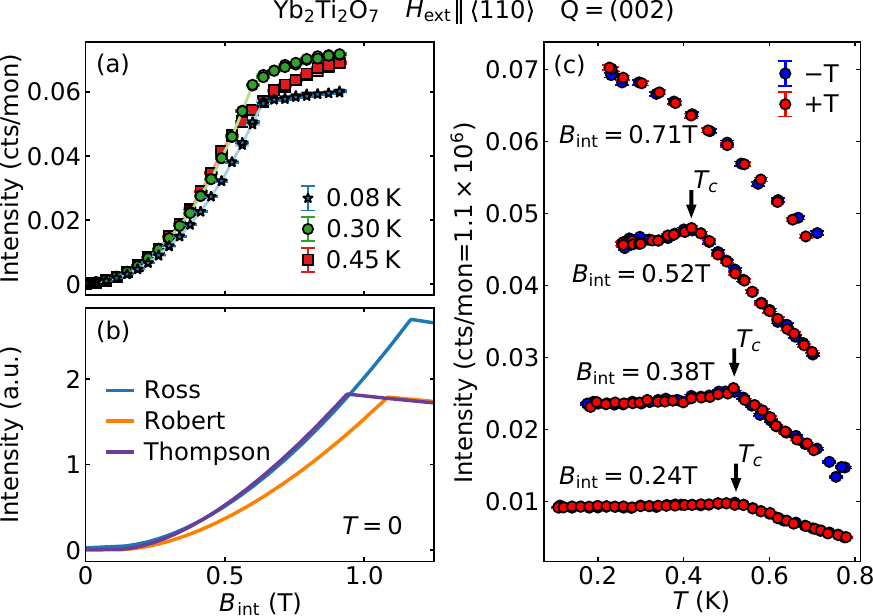}
	\caption{Neutron scattering of the (002) peak in \ce{Yb2Ti2O7} for magnetic fields applied along the $\langle 110 \rangle$ direction. (a) Field dependence of (002) at different temperatures, showing a quadratic dependence at low fields and a clear upper critical field. (b) Theoretical (002) scattering calculated with mean-field theory using the Ross, Robert, and Thompson Hamiltonians \cite{2011_Ross_PhysRevX,2015_Robert_PhysRevB,2017_Thompson_PhysRevLett}. (c) Temperature dependent scattering at different applied fields. No hysteresis is visible.  Error bars represent one standard deviation.} 
	\label{fig:figure_4_002_nscattering}
\end{figure}

Qualitatively, the orientation dependence of the magnetic phase diagram of \ce{Yb2Ti2O7}, shown in \fig\,\ref{fig:figure_1_phasediagram} for low temperatures, is consistent with the behavior of a cubic ferromagnet, \ie where cubic magnetocrystalline anisotropy selects six ground states with magnetization along $\langle 100 \rangle$ \cite{1937_vanVleck_PhysRev}. In $\langle 111 \rangle$ and $\langle 110 \rangle$ fields, however, the phase diagram exhibits a highly unusual field dependence, wherein an applied magnetic field initially increases the ordering temperature and then suppresses it at higher fields, which results in a reentrant phase diagram (\fig\ref{fig:figure_1_phasediagram}(a) and (b)). For fields along $\langle 100 \rangle$, the high-field phase boundary collapses and the system enters a field polarized state for small applied fields, which is qualitatively distinct from the other field directions (\fig\ref{fig:figure_1_phasediagram}(c)). This orientation dependence of the magnetic phase diagram was determined by measurements of the temperature and field dependent magnetization (\fig\ref{fig:figure_2_MvsT_MvsB}), heat capacity (\fig\ref{fig:figure_3_HC_ACDR}(a)), susceptibility (\fig\ref{fig:figure_3_HC_ACDR}(b)), and neutron diffraction (\fig\ref{fig:figure_4_002_nscattering}). Hysteretic effects observed under field and temperature sweeps are indicated by means of blue and red shading, respectively. While the $\langle 111 \rangle$ data was reported in a previous study \cite{2017_Scheie_PhysRevLett}, the two other directions, which are essential for the conclusions of our study, are reported here for the first time.

The $\langle 111 \rangle$ and $\langle 110 \rangle$ phase diagrams have nearly the same upper critical field (\SI{0.63}{\tesla} and \SI{0.57}{\tesla}), but the reentrance for fields along $\langle 110 \rangle$ is even more dramatic: the highest $\langle 110 \rangle$ $T_{\mathrm{C}}$ at $B_{\mathrm{int}}\,=\,0.30\,$\si{T} is \SI{540}{mK}, which is a 100$\,$\% increase above zero field $T_{\mathrm{C}}\,=\,270\,$\si{mK}. ($\langle 111 \rangle$, meanwhile, has a 55\% increase.) 

Applying the field along $\langle 100 \rangle$ polarizes the system already for small applied field (as can be seen in the temperature dependence of the magnetization in \fig\ref{fig:figure_2_MvsT_MvsB}(C)),  so there is no high-field phase boundary (as shown by the field dependence of the magnetization in \fig\ref{fig:figure_2_MvsT_MvsB}(F)). In the absence of magnetic field, the ground state of \ce{Yb2Ti2O7} is ferromagnetic with magnetization spontaneously breaking the six-fold degenerate $\langle 100 \rangle$ directions. For a field applied along $\langle 100 \rangle$, there is no spontaneous symmetry breaking, hence no phase transition \cite{2017_Thompson_PhysRevLett}. For a first order zero field transition like in \ce{Yb2Ti2O7}, however, the transition should survive for small, but finite fields \cite{2017_Thompson_PhysRevLett}, which also is fully consistent with the data.

Going from the $\langle 111 \rangle$ via $\langle 110 \rangle$ to the $\langle 100 \rangle$ direction, the magnetization shows an increase in the spontaneous magnetic moment (\fig\ref{fig:figure_2_MvsT_MvsB}(D-F)) consistent with the behavior of a cubic ferromagnet. The coercive field in the ferromagnetic regime of \ce{Yb2Ti2O7} is vanishingly small, which indicates extremely weak domain wall pinning (\fig\ref{fig:figure_2_MvsT_MvsB}(D-F) and supplementary online information \fig\ref{fig:figure_8_sup_MvsB_hysteresis_111_coercivity}).

We previously argued, through comparison to classical simulations, that the reentrant nature of the $\langle 111 \rangle$ phase boundary is due to quantum fluctuations suppressing the ferromagnetic order \cite{2017_Scheie_PhysRevLett}. Exact diagonalization calculations, using ground state and finite temperature methods, support this hypothesis \cite{2017_Changlani_arxiv}. We anticipate the same explanation holds for the even more extreme reentrance observed for fields along the $\langle 110 \rangle$  direction.

While previous $\langle 111 \rangle$ measurements indicated a first order phase boundary \cite{2017_Scheie_PhysRevLett}, the  $\langle 110 \rangle$ data provides evidence of a second order (continuous) phase boundary---at least for the upper critical field---in three ways. First, the heat capacity (\fig\ref{fig:figure_3_HC_ACDR}(a)) shows a lambda-like anomaly (in contrast to the symmetric peaks seen in the $\langle 111 \rangle$ direction \cite{2017_Scheie_PhysRevLett}), and this is a signature of a second order phase transition \cite{2000_Chaikin_Book}. Second, the susceptibility (\fig\ref{fig:figure_3_HC_ACDR}(b)) has a step-like feature, which---because susceptibility is related to the second field derivative of free energy---indicates a discontinuity in the second derivative of free energy and thus a second order phase transition. Third, magnetization (\fig\ref{fig:figure_2_MvsT_MvsB}) and neutron scattering (\fig\ref{fig:figure_4_002_nscattering}) field and temperature sweeps detect no hysteresis at the high-field phase boundary, even for faster magnetization field sweep rates of \SI{60}{mT/min}, which suggests a continuous phase transition.

There are three caveats to this second-order boundary hypothesis, all centered on observing hysteresis: (i) there is noticeable hysteresis in the high-field phase boundary of the $\langle 110 \rangle$ susceptibility measurements (\fig\ref{fig:figure_3_HC_ACDR}(b)), (ii) substantial hysteresis is observed at the lower phase boundary in $\langle 110 \rangle$ magnetization data (\fig\ref{fig:figure_2_MvsT_MvsB}(E)), and (iii) hysteresis is observed in the temperature sweeps of $\langle 110 \rangle$ magnetization (\fig\ref{fig:figure_2_MvsT_MvsB}(B3) and (B4)). Typically, hysteresis is a signature of a first order transition via nucleation and domain growth. That very much seems to be the case for the lower field part of the phase boundary (where $T_{\mathrm{C}}$ increases with field), especially from (iii): the hysteresis in $M$ vs $T$ (\fig\ref{fig:figure_2_MvsT_MvsB}(B)).

The magneto-caloric effect, however, offers an alternative explanation for the hysteresis observed at the high-field phase boundary. Examining the susceptibility data in \fig\ref{fig:figure_3_HC_ACDR}(b) closely, the sweep under increasing field (solid lines) displays phase transitions at lower fields than the sweep under decreasing field (dashed lines). This is the opposite of what may be expected for a first order phase transition. In this case, there should be a \textit{delay} in the onset of the phase transition, not a speeding-up. Instead, what seems to occur is that the sample, when it crosses the lower phase boundary into the ordered phase, experiences a large magneto-caloric effect due to the release of entropy. This causes the sample to heat such that it crosses the high-field phase boundary at a slightly higher temperature than it does when cooling down again---leading to an apparent hysteresis in susceptibility proportional to the slope of the phase boundary. In \fig\ref{fig:figure_3_HC_ACDR}(b) and (d), this is illustrated for the susceptibility data taken at $0.5\,$K. This interpretation was confirmed by fast field-sweeps measuring magnetization with reduced thermal coupling to the refrigerator: at \SI{400}{mK}, no transition was observed on increasing field but a transition was observed while decreasing field (the sample heated so much as to avoid the phase boundary entirely). The magneto-caloric effect at a second order phase transition is consistent with all of these observations.

It is difficult to prove a transition to be first or second order based on our experiments alone. However, our data are entirely consistent with a second order transition for the $\langle 110 \rangle$ phase boundary while there would be inconsistencies for a first order transition.

\subsection*{Theory}

We now discuss the theoretical framework needed for a full account of our observations. Coarse-grained theoretical calculations confirm the order of the phase transitions observed in the low temperature magnetization, and show that \ce{Yb2Ti2O7} behaves qualitatively like a cubic ferromagnet such as iron. We also show that classical mean-field theory cannot account for the high-field phase boundaries, which indicates that the high-field phase boundary is subject to collective or quantum effects. 

\begin{figure*}[htbp]
	\centering
	\includegraphics[width=\textwidth]{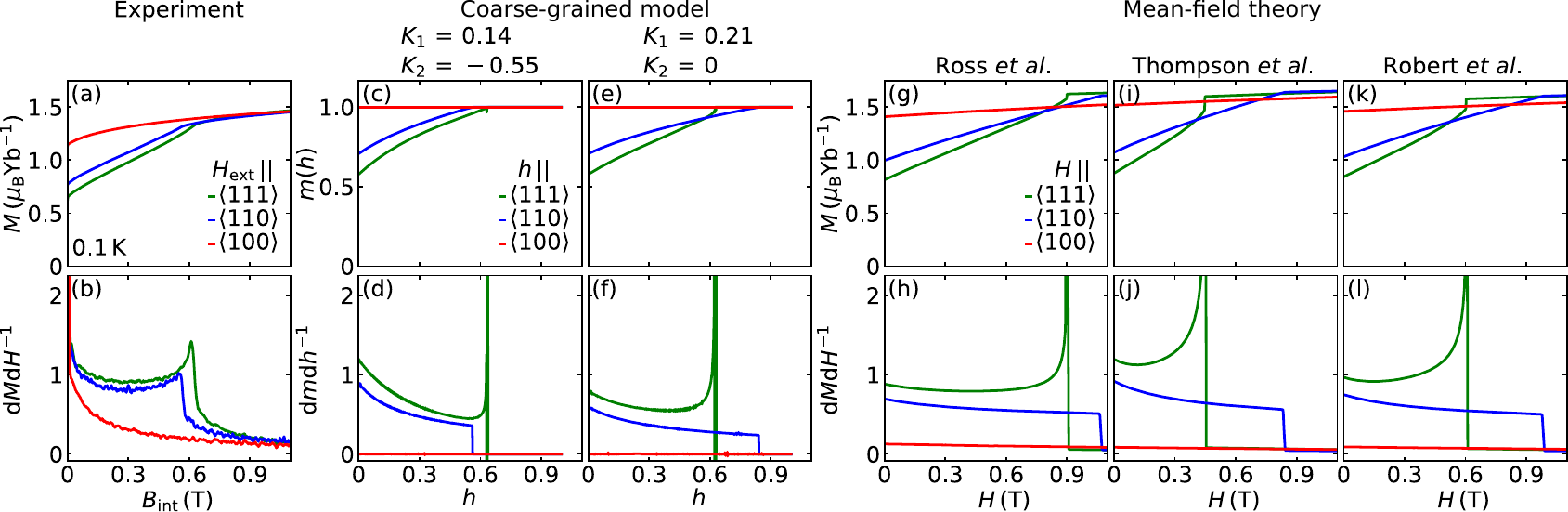}
	\caption{Experimental data (a,b) and calculated magnetization via the coarse-grained model (c-f) and mean-field theory (g-l) for \ce{Yb2Ti2O7}.
	(a) Magnetic field dependence of the magnetization of \ce{Yb2Ti2O7} at \SI{0.1}{\kelvin} for the $\langle 111 \rangle$, $\langle 110 \rangle$, and $\langle 100 \rangle$ direction.
	(b) Differential susceptibility $\mathrm{d}M/\mathrm{d}H$ calculated from the magnetization data.
	(c-f) Magnetization versus field as obtained from the coarse-grained model and differential susceptibility.
	(g-l) Mean-field calculation at $T\,=\,0$, taking into account a cubic anisotropy and the Zeeman field. Calculations were performed for the exchange parameters from Ross \etal\cite{2011_Ross_PhysRevX} (g,h), Thompson \etal\cite{2017_Thompson_PhysRevLett} (i,j), and Robert \etal\cite{2015_Robert_PhysRevB} (k,l).
 	(h,j,l) Susceptibility calculated from the theoretical model of the magnetization shown in panels (g,i,k).}
	\label{fig:figure_5_ExpvsTheo_allinone}
\end{figure*}

\subsubsection*{Coarse-grained model}

We consider a coarse-grained model to describe the magnetization for magnetic fields applied in the three main symmetry directions. A coarse-grained picture is based on the uniform magnetization associated with the sum of the four adjacent spins on a tetrahedron. The six ground states with a canted ferromagnetic order thus yield the uniform magnetization pointing along one of the six $\langle 100 \rangle$ directions. When the sample is magnetized by domain selection only (at the largest applied field where the internal magnetic field is zero), the projection of the magnetization to $\langle 100 \rangle$, $\langle 110 \rangle$, and $\langle 111 \rangle$ directions has a ratio of $1\,:\,1/\sqrt{2}\,:\,1/\sqrt{3}$, indicating $\langle 100 \rangle$ as the easy-axis. The experimentally obtained ratios of spontaneous moment and fields match the theoretical prediction well as summarized in \tab\ref{tab:theory-ratios}.

\begin{table}[htbp]
\centering
\caption{Spontaneous magnetic moment from \ce{Yb2Ti2O7} as extrapolated for zero field from the initial field dependence along $\langle 100 \rangle$, $\langle 110 \rangle$, and $\langle 111 \rangle$. The ratios between the spontaneous moments are in excellent agreement with theoretical predictions.}
\label{tab:theory-ratios}
\begin{tabular}{lccc}
\hline 
$H_{\mathrm{ext}}\,||\quad$				& $M_0\,$($\mu_{\mathrm{B}}$\ce{Yb^{-1}})$\quad$ & $M_0/M_{0,100}\quad$  & ratio theory     \\ \hline
$\langle 100 \rangle$                   & $1.197(14)$                           & $1$              & $1$        \\
$\langle 110 \rangle$                   & $0.828(3)$                           & $0.692(13)$           & $1/\sqrt{2}\,\approx\,0.707$     \\
$\langle 111 \rangle$                   & $0.678(6)$                           & $0.566(12)$           & $1/\sqrt{3}\,\approx\,0.577$     \\ \hline
\end{tabular}
\end{table}

The cubic anisotropy is minimized with a six-fold degeneracy for magnetization along $\{ \pm \vec{e}_i \}$, \ie $\vec{e}_i \,=\, \hat{\vec{x}}, \hat{\vec{y}}, \hat{\vec{z}}$ in the global frame. Ignoring higher-order terms, the potential energy for the magnetization represented by a unit vector $\vec{m}$ is 
\begin{equation}
\label{eq:anisotropy-zeeman-potential}
U = -K_1 \sum_i (\vec{m} \cdot \vec{e_i})^4 - K_2 \prod_i (\vec{m} \cdot \vec{e_i})^2 - \vec{h} \cdot \vec{m}.
\end{equation}
Both $K_1$ and $K_2$ are cubic anisotropy terms: a positive $K_1$ makes the minimum energy direction along $\langle 100 \rangle$, and positive $K_2$ makes the minimum energy along $\langle 111 \rangle$, but with a different angular dependence than for negative $K_1$. The minimization of the cubic anisotropy is reported in the online supplemental information.

From the experimental measurement $h_c^{\langle 110 \rangle} \,=\, 0.57$ and $h_c^{\langle 111 \rangle} \,=\, 0.63$, we derive the parameters in the anisotropy model to be $K_1 \,=\, 0.14$ and $K_2 \,=\, -0.55$. Minimizing the potential energy (see supplementary online information \eq{\ref{eq-sub:anisotropy-zeeman-potential}}) under the constraint $\vec{m}^2 \,=\, 1$ gives the magnetization response to magnetic fields (\fig\ref{fig:figure_5_ExpvsTheo_allinone}(c)). 

As a comparison, we set $K_2 \,=\, 0$ and look at the lower-order cubic anisotropy with $K_1 = 0.21$ to reproduce the measured transition field in $\langle 111 \rangle$ direction $h_c^{\langle 111 \rangle} \,=\, 0.63$. The obtained magnetization curve (\fig\ref{fig:figure_5_ExpvsTheo_allinone}(e)) appears to be closer to the result from the classical mean-field calculations (\fig\ref{fig:figure_5_ExpvsTheo_allinone}(g,i,k)) than the actual measurements (\fig\ref{fig:figure_5_ExpvsTheo_allinone}(a)).

The effect of the two anisotropy terms can be seen in a simple evaluation of the potential energy in zero field for $\vec{m} \,=\, (1,0,0), (1,1,0)/\sqrt{2}$, and  $(1,1,1)/\sqrt{3}$, respectively, yielding $-K_1$, $-K_1/2$, and $-K_1/3-K_2/27$. For easy axis along $\langle 100 \rangle$, $K_1>0$, a negative $K_2$ makes $\langle 111 \rangle$ an even harder axis.
Even more interesting is that applying Landau theory \cite{2013_Landau_book} to this simple coarse-grained model predicts a second order phase boundary for a $\langle 110 \rangle$ field and a first order phase boundary for a $\langle 111 \rangle$ field (see supplemental online information), consistent with our experimental observations.
This exercise in coarse-grained modeling shows that the base temperature magnetization and the order of the phase boundaries can be understood as the effects of cubic anisotropy.

\subsubsection*{Classical mean-field theory}

To better understand the behavior of individual spins, we apply classical mean-field calculations to the Hamiltonian
\begin{align}
	\label{eq:Hamiltonian}
    \mathcal{H} \,=\, \frac{1}{2} \sum_{ij} J_{ij}^{\mu\nu} S_{i}^{\mu} S_{j}^{\nu} - \mu_{\mathrm{B}} H^{\mu} \sum_{i} g_{i}^{\mu\nu} S_{i}^{\nu},
\end{align}
where $J_{ij}^{\mu\nu}$ is the matrix of exchange couplings and $g_{i}^{\mu\nu}$ the g-tensor (see \reftaken\cite{2011_Ross_PhysRevX} for notation),
using experimentally determined exchange parameters from literature \cite{2011_Ross_PhysRevX,2017_Thompson_PhysRevLett,2015_Robert_PhysRevB} to describe the magnetization (\fig\ref{fig:figure_5_ExpvsTheo_allinone}(g-l)). For each parameter set we find the classical ${\bf Q}\,=\,0$ state that minimizes the Hamiltonian in \eq{\ref{eq:Hamiltonian}} and extract the field dependence of the magnetization projected along the field direction. This model accurately describes the field-dependent neutron scattering (\fig\ref{fig:figure_4_002_nscattering}). However, it predicts a lower critical field for field along $\langle 111 \rangle$ than for field along $\langle 110 \rangle$, which is opposite to the experimental result.

The field-dependent spin configurations from mean-field calculations allows us to calculate the neutron scattered intensity, which agrees well with our experimental results (\fig\ref{fig:figure_4_002_nscattering}(a,b)) and shows a non-collinear spin structure in \ce{Yb2Ti2O7}. In general, with field-dependent magnetic Bragg intensities it should be possible to track the magnetic structure as a function of magnetic field. Unfortunately, the majority of peaks exhibit field-dependent extinction which is typical for ferromagnets \cite{1989_Baruchel_PhysicaBCondensedMatter} (see online supplementary information) and complicates interpretation of the experimental data. Magnetic scattering with minimal extinction is only observed on the weakest Bragg peak, (002), which still affords a view into the spin correlations as a function of magnetic field. The magnetic neutron structure factor for (002) on the pyrochlore lattice is given by
\begin{equation}
\label{eq:002scattering}
\begin{array}{ll}
    S({\bf Q}=(002)) = & 6\big[{\bf S}_{1}^{2}+{\bf S}_{2}^{2} +{\bf S}_{3}^{2} +{\bf S}_{4}^{2} \\
    & + 2({\bf S}_{1}\cdot{\bf S}_{2} -{\bf S}_{2}\cdot{\bf S}_{3} -{\bf S}_{1}\cdot{\bf S}_{3}
    \\
     & \enskip \enskip -{\bf S}_{1}\cdot{\bf S}_{4}-{\bf S}_{2}\cdot{\bf S}_{4}+{\bf S}_{3}\cdot{\bf S}_{4} ) \big],
\end{array}
\end{equation}
where spins ${\bf S}_{1}$, ${\bf S}_{2}$, ${\bf S}_{3}$, and ${\bf S}_{4}$ are the four spins on a tetrahedron. As is evident from this equation, a fully polarized spin state (${\bf S}_{1}={\bf S}_{2}={\bf S}_{3}={\bf S}_{4}$) has zero neutron intensity. Thus, (002) intensity is a direct measure of the non-collinearity of the spin structure.

\begin{figure}[htbp]
	\centering
	\includegraphics[width=\columnwidth]{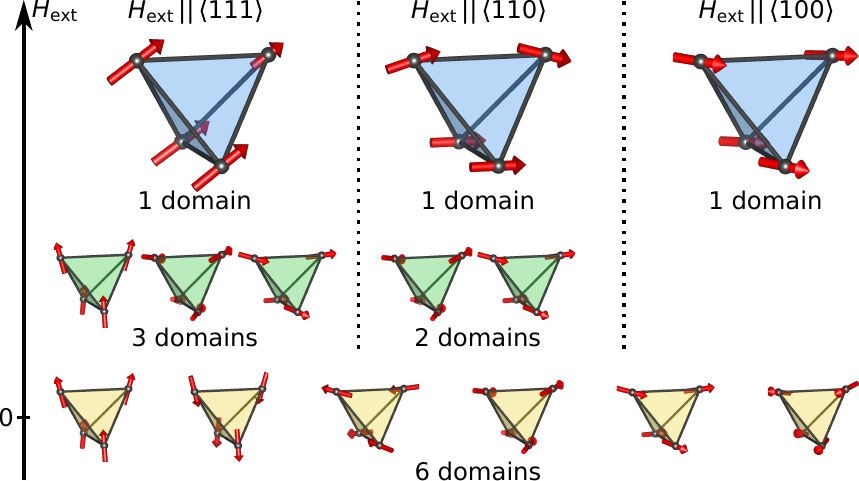}
	\caption{Field dependent magnetic structure of \ce{Yb2Ti2O7} for applied fields along $\langle 111 \rangle$, $\langle 110 \rangle$, and $\langle 100 \rangle$. In small fields, out of the six domains the system selects three and two domains for field along $\langle 111 \rangle$ and $\langle 110 \rangle$, respectively. In higher fields, the spins enter a polarized state where the spins either lie in or are canted towards their easy-plane defined by the local $\langle 111 \rangle$ axis. For field along $\langle 100 \rangle$, application of a magnetic field immediately stabilizes the configuration shown in blue shading.}
	\label{fig:figure_6_spin_structures}
\end{figure}

This means that the increase in $(002)$ intensity with $\langle 110 \rangle$ field up to an external field of \SI{1}{T} signifies that the region above the upper critical field is not uniformly polarized. This behavior is reproduced by the mean-field simulations (\fig\ref{fig:figure_4_002_nscattering}(b)) and shows spins which either lie in or are canted towards their easy-planes defined by the local $\langle 111 \rangle$ axis \cite{2015_Gaudet_PhysRevB}, as depicted in \fig\ref{fig:figure_6_spin_structures}.

Despite the success of classical mean-field theory in qualitatively describing the field evolution of the spin structure, it incorrectly predicts that the boundary of the high-field phase for fields along $\langle 111 \rangle$ is lower than for fields along $\langle 110 \rangle$. Experimentally, the opposite is observed ((\fig\ref{fig:figure_5_ExpvsTheo_allinone}(g-l)). The origin of this discrepancy is beyond the analysis presented so far and must be left for the future.

\section{Conclusions}

Our observations clarify several important issues surrounding \ce{Yb2Ti2O7} and highlight its exceptional properties. We demonstrate that the magnetization in \ce{Yb2Ti2O7} is characteristic of a cubic ferromagnet where the low-field behavior is governed by simple magnetic domain selection. We also find that the order of the phase boundary for $\langle 110 \rangle$ and $\langle 111 \rangle$ as well as the lack of phase boundary for $\langle 100 \rangle$ are consistent with the predictions of Landau theory for a cubic ferromagnet. However, the ratio of upper critical fields for $\langle 110 \rangle$ and $\langle 111 \rangle$ is inconsistent with mean-field theory, suggesting the presence of strong correlations. Inferred from elastic neutron scattering, the field dependent magnetic structure shows that the field-polarized phase is not collinear but has the spins canted towards the easy-plane orthogonal to the local $\langle111\rangle$ pyrochlore axes.  We also reveal a dramatic reentrant phase diagram for field along $\langle 110 \rangle$ as previously reported for $\langle 111 \rangle$, suggesting that the low-field finite temperature regime is a state where highly unconventional correlations dominate \cite{2017_Changlani_arxiv}.

While \ce{Yb2Ti2O7} appears to be a deceptively simple cubic ferromagnet at low temperatures, the reentrant phase diagram and the reversed anisotropy of the upper critical fields are clear experimental findings that cannot be accounted for by classical microscopic theory. Instead they indicate that the paramagnetic state near the phase boundary for $T\,=\,0$ and $B\,=\,0$ are theoretically challenging regimes where strong correlations prevail. The importance of this ferromagnetic state as a point of reference in the exploration of unconventional correlations is underscored by the recent observation of meso-scale antiferromagnetic textures as well as ferro- and antiferromagnetic spin waves \cite{2019_Scheie_ArXiv}.

\section*{Methods}

\subsection*{Magnetization}

The magnetization of \ce{Yb2Ti2O7} was measured by means of a bespoke vibrating coil magnetometer (VCM) combined with a TL400 Oxford Instruments top-loading dilution refrigerator \cite{NIST_disclaimer}, as described in \reftaken \cite{2010_Legl_RevSciInstrum,2012_Krey_PhysRevLett,2012_Legl_PhysRevLett}. Data were recorded at temperatures down to \SI{0.028}{\kelvin} under magnetic fields up to \SI{5}{\tesla} at a low excitation frequency of \SI{19}{\hertz} and a small excitation amplitude of $\sim\,$\SI{0.5}{\milli\meter} (the measurement protocols are described in the supplementary online information). The sample temperature was measured with a ruthenium oxide sensor mounted next to the sample and additionally monitored with a calibrated Lakeshore ruthenium oxide temperature sensor attached to the mixing chamber in the zero-field region. 

For our magnetization measurements a single crystal was cut from an ingot and carefully ground and polished into a spherical shape with a diameter of $\sim\SI{4.7}{\milli\meter}$. The high-quality stoichiometric single crystal of \ce{Yb2Ti2O7} was grown with the traveling solvent floating zone method as described by Arpino \etal \cite{2017_Arpino_PhysRevB}. Samples from the same crystal were previously used in the study of Scheie \etal \cite{2017_Scheie_PhysRevLett}. A spherical sample shape was chosen to minimize inhomogeneities of the demagnetizing fields, permitting straightforward computation of the internal field values. To suspend the sample in the VCM it was glued with GE varnish into an oxygen-free copper sample holder composed of two matching sections accurately fitting the size of the sphere (further information on sample mounting are in the supplementary online information). The data are plotted in \fig\ref{fig:figure_2_MvsT_MvsB}.

\subsection*{Heat Capacity}

The heat capacity of \ce{Yb2Ti2O7} was measured using a Quantum Design PPMS \cite{NIST_disclaimer} using LongHCPulse \cite{2018_Scheie_JLowTempPhys}. Over four days, the temperature dependent specific heat was measured at 18 magnetic fields between $0$ and \SI{1}{\tesla}  with the long pulse method, and at one magnetic field using the short pulse method. This measurement was performed on a \SI{1.04}{mg}, $\SI{1.1}{mm} \times \SI{0.6}{mm} \times \SI{0.2}{mm}$  prism of \ce{Yb2Ti2O7} with a demagnetization factor of $0.59$ for fields along the shortest dimension which was the $\langle 110 \rangle$ axis. This sample was cut from the same crystal as the heat capacity sample in \reftaken \cite{2017_Scheie_PhysRevLett}. The results are shown in \fig\ref{fig:figure_3_HC_ACDR}(a).

\subsection*{AC Susceptibility}

The AC susceptibility of \ce{Yb2Ti2O7} was measured using a Quantum Design PPMS with an AC susceptibility dilution refrigerator insert \cite{NIST_disclaimer}. The measurement was performed on a \SI{59}{mg}, $\SI{2.5}{mm} \times \SI{2.0}{mm} \times \SI{1.6}{mm}$ cuboid with a demagnetization factor of $0.406$ for fields along the shortest dimension which was the $\langle 110 \rangle$ axis. The sample was glued to a sapphire rod with GE varnish to ensure good thermal connection. This sample was cut from a different \ce{Yb2Ti2O7} crystal than the heat capacity and magnetization samples, still grown by the same method as described in \reftaken \cite{2017_Arpino_PhysRevB}. We measured the real ($\chi '$) and imaginary ($\chi ''$) susceptibility as a function of field at different temperatures, sweeping the $\langle 110 \rangle$ magnetic field at \SI{60}{mT/min} from  $0\,\rightarrow\,$\SI{+1}{\tesla}$\,\rightarrow\,$\SI{-1}{\tesla}$\,\rightarrow\,$\SI{0}{\tesla} and measuring susceptibility with \SI{1}{kHz} and an AC field amplitude of $1\,$Oe. Tests with different frequencies and sweep rates revealed that the anomalies at the upper critical field are not frequency or sweep-rate dependent. The data are plotted in \fig\ref{fig:figure_3_HC_ACDR}(b-c).

\subsection*{Neutron Scattering}

The elastic neutron scattering from \ce{Yb2Ti2O7} was measured using the SPINS triple axis spectrometer at the NCNR. We used a spherical sample (the same sphere as in \reftaken\cite{2017_Scheie_PhysRevLett}) with the $[1 \bar 1 0]$ direction along a vertical magnetic field and mounted in a dilution refrigerator. We collected field and temperature dependent elastic scattering on the $(111)$, $(002)$, $(220)$, $(113)$, $(222)$, and $(004)$ Bragg peaks using \SI{4.5}{meV} neutrons. The collimations were guide$\,$-$\,80^{\prime}\,$-$\,80^{\prime}\,$-$\,$open, and Be filters were used before and after the sample.
Unfortunately, significant field-dependent extinction precluded the field-dependent intensities from being compared to theory (see supplementary online information for details) with the exception of the (002) peak---the only magnetic peak with zero nuclear intensity---which is shown in \fig\ref{fig:figure_4_002_nscattering}. The data from other peaks are shown in the supplementary online information.
The phase transition is clearly visible as a kink in the data, and can be tracked as a function of magnetic field.
The scattering data were acquired after centering the detector on the Bragg peak using rocking and $\theta-2\theta$ scans, but there is some imprecision in doing this so that some scans have slightly attenuated intensities compared to others (\eg the \SI{80}{mK} field scan in \fig\ref{fig:figure_4_002_nscattering} should have higher intensity).

\subsection*{Acknowledgments}
The authors wish to thank A. Bauer, C. Castelnovo, M. Kleinhans, S. Mayr, R. Moessner, N. Shannon, and A. Wendl for support and stimulating discussions.
Financial support through DFG TRR80 (107745057), DFG Excellence Strategy – EXC-2111 (390814868), ERC Advanced Grant ExQuiSid (788031), and BMBF Project RESEDA-Plus (05K16WO6) is gratefully acknowledged. SS and CD acknowledge financial support through the TUM Graduate School.
This work was supported as part of the Institute for Quantum Matter, an Energy Frontier Research Center funded by the U.S. Department of Energy, Office of Science, Basic Energy Sciences under Award No. DE-SC0019331. AS and CB were supported through the Gordon and Betty Moore foundation under the EPIQS program GBMF-4532. The dilution refrigerator used for magnetization and AC susceptibility measurements was funded by the NSF through DMR-0821005.
Use of the NCNR facility was supported in part by the National Science Foundation under Agreement No. DMR-1508249.
Travel for experimental work at TUM by AS was funded by a QuantEmX grant from ICAM and the Gordon and Betty Moore Foundation through Grant GBMF-5305.
HJC was supported by start up funds from Florida State University and the National High Magnetic Field Laboratory. The National High Magnetic Field Laboratory is supported by the National Science Foundation through NSF/DMR-1644779 and the state of Florida.

\bibliography{library.bib}

\begin{thebibliography}{49}
\bibitem [{NIS()}]{NIST_disclaimer}%
  \BibitemOpen
  \href@noop {} {\enquote {\bibinfo {title} {Certain commercial instruments are
  identified in this paper to foster understanding. such identification does
  not imply recommendation or endorsement by the national institute of
  standards and technology, nor does it imply that the instruments identified
  are necessarily the best available for the purpose.}}\ }\BibitemShut
  {NoStop}%
\end{thebibliography}

\clearpage

\section*{Supplemental Materials}

\subsection*{Magnetization}

\begin{figure}[htbp]
	\centering
	\includegraphics[width=\columnwidth]{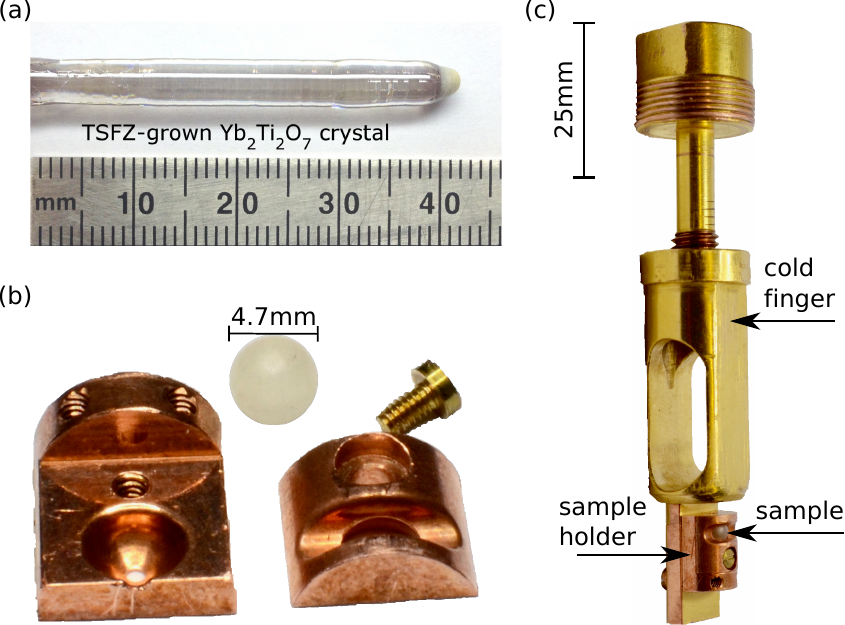}
	\caption{(a) Stoichiometric, pure, and colourless \ce{Yb2Ti2O7} single crystal grown by the traveling solvent floating zone (TSFZ) technique (image taken from \reftaken\cite{2017_Arpino_PhysRevB}).
	(b) spherical sample ground from the stoichiometric single crystal and the oxygen-free \ce{Cu} sample holder composed of two matching sections fitting accurately the size of the sphere.
	(c) sample holder mounted on the cold finger which is then bolted to the \ce{Cu} tail attached to the mixing chamber of the dilution refrigerator.}
	\label{fig:figure_7_sup_sample}
\end{figure}

The sample, sample holder, and cold finger for the magnetization measurements are shown in \fig\ref{fig:figure_7_sup_sample}. The sample holder with the sample mounted was firmly bolted into a \ce{Cu} tail attached to the mixing chamber of the dilution refrigerator. This provided excellent thermal anchoring of the sample across the entire surface of the sphere during all measurements, while keeping its position rigidly fixed mechanically without exerting significant stress.

The magnetization data were recorded following well-defined field and temperature histories. Concerning the temperature dependence three procedures were used: (i) After cooling at zero magnetic field from $\sim\,$\SI{1}{\kelvin}, the magnetic field was applied at base temperature and data collected while heating continuously at a rate of \SI{5}{\milli\kelvin\per\minute}. This is referred to as zero-field-cooled / field-heated (zfc-fh). (ii) Data were recorded while cooling in the same unchanged applied magnetic field. This is referred to as field-cooled (fc) (iii) After initially cooling in the applied magnetic field, data were recorded while heating continuously at a rate of \SI{5}{\milli\kelvin\per\minute} in the same unchanged magnetic field. These data are referred to as field-cooled / field-heated (fc-fh). Similarly, isotherms were collected in one of the following three different field sweeps: (iv) After zero-field-cooling a sweep from $0\,\rightarrow\,$\SI{+1}{\tesla}, denoted (A1). (v) A field sweep starting at a high field, notably from \SI{+1}{\tesla}$\,\rightarrow\,$\SI{-1}{\tesla}, denoted (A2). (vi) A related field sweep from \SI{-1}{\tesla}$\,\rightarrow\,$\SI{+1}{\tesla}, denoted (A3). For temperatures above \SI{0.05}{\kelvin} all data were recorded while sweeping the field continuously at \SI{15}{\milli\tesla\per\minute}, whereas the measurement at \SI{0.022}{\kelvin}, the lowest temperature accessible, was carried out at a continuous sweep rate of \SI{1.5}{\milli\tesla\per\minute} to minimize eddy current heating of the \ce{Cu} tail.

To isolate the signal from the sample, data was also acquired for the empty sample holder and subtracted from the data acquired with the sample in place. The signal of the empty sample holder was found to be small with a highly reproducible field dependence and negligible temperature dependence. The sample signal was calibrated quantitatively at \SIlist{2;3}{\kelvin} against the magnetization measured in a Quantum Design physical properties measurement system determined also at \SIlist{2;3}{\kelvin}, as well as a \ce{Ni} standard measured separately in the VCM \cite{2010_Legl_RevSciInstrum}.

\begin{figure}[htbp]
	\centering
	\includegraphics[width=\columnwidth]{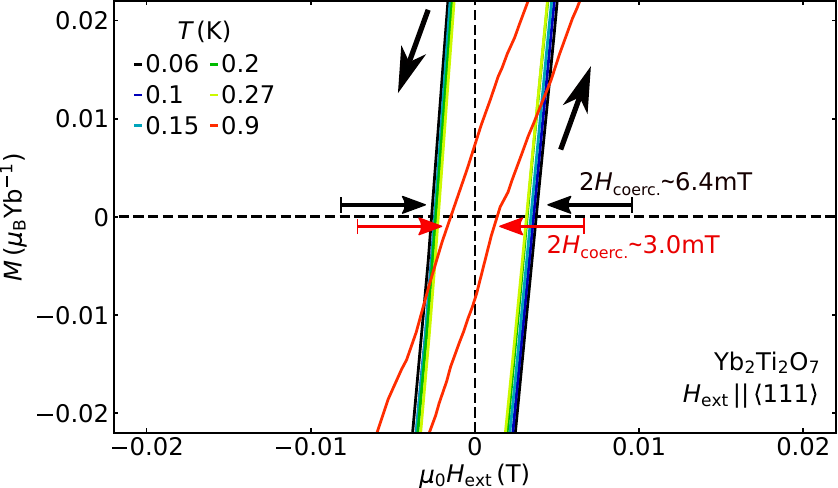}
	\caption{Magnetic hysteresis in \ce{Yb2Ti2O7} for temperatures $T\,\leq\,T_{\mathrm{C}}$ (\SI{0.06}{K}, \SI{0.10}{K}, \SI{0.15}{K}, \SI{0.20}{K}, and \SI{0.27}{K}), and for $T\,=\,0.9\,\si{K}$ in the paramagnetic regime. The coercive field $H_{\mathrm{coerc.}}$ in the paramagnetic regime is finite due to instrumental resolution around $H\,=\,0$. The coercive field in the ferromagnetic regime of \ce{Yb2Ti2O7} is vanishingly small, suggesting that magnetic domain walls in the ordered state move freely in response to an applied magnetic field.}
	\label{fig:figure_8_sup_MvsB_hysteresis_111_coercivity}
\end{figure}

\fig\ref{fig:figure_8_sup_MvsB_hysteresis_111_coercivity} shows the magnetic hysteresis in \ce{Yb2Ti2O7}. The coercive field in the ferromagnetic regime of \ce{Yb2Ti2O7} is vanishingly small, suggesting that magnetic domain walls in the ordered state can move almost freely upon applying a magnetic field.

\subsection*{Neutron Scattering}

\begin{figure*}[htbp]
	\centering
	\includegraphics[width=\textwidth]{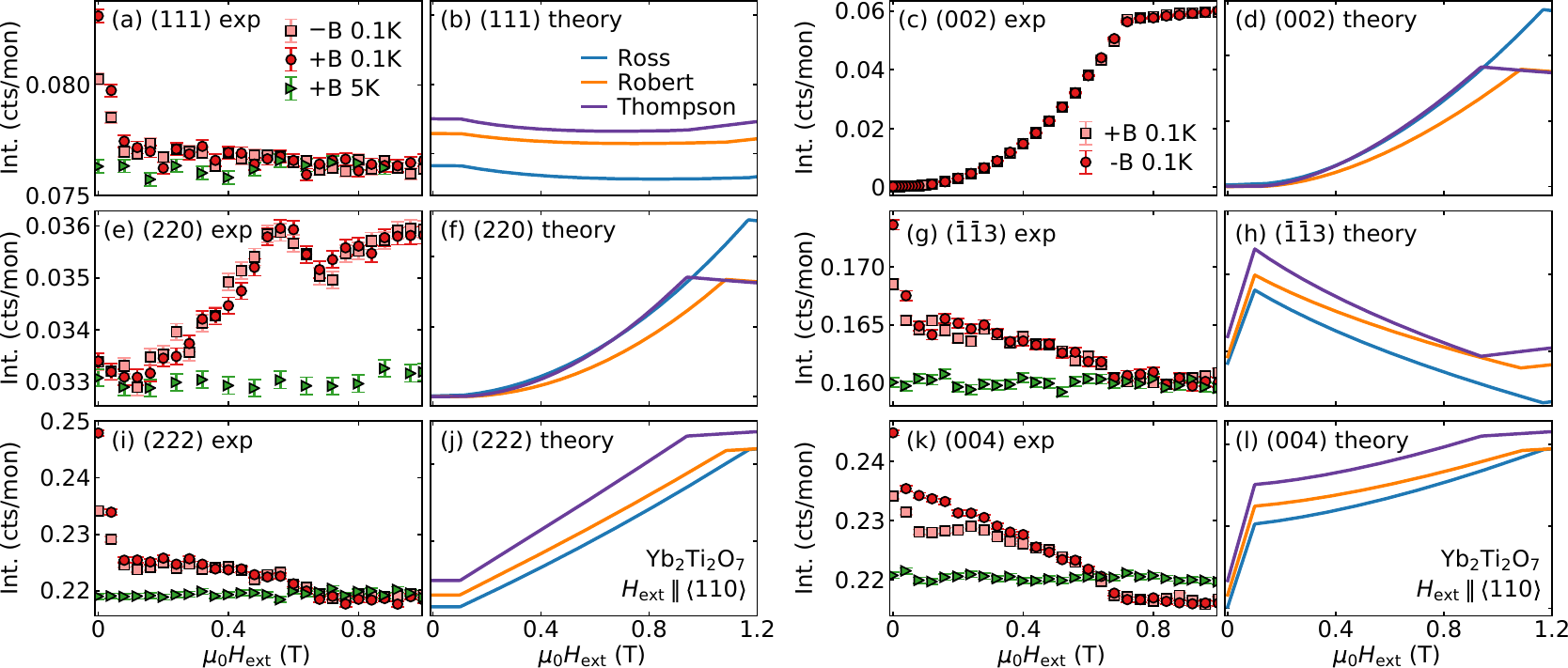}
	\caption{Field dependent neutron scattering from \ce{Yb2Ti2O7} at \SI{0.1}{K} and \SI{5}{K} compared to theoretical calculated intensity from mean-field simulations. For strong Bragg peaks the data does not match the simulations because of field-dependent extinction, but for the weak Bragg peaks (002) and (220) the theory does match. Error bars represent one standard deviation.}
	\label{fig:figure_9_sup_neutron_field_dependence_vs_theory}
\end{figure*}

The field dependent elastic neutron scattering data are compared to mean-field simulations based on the Ross \cite{2011_Ross_PhysRevX}, Robert \cite{2015_Robert_PhysRevB}, and Thompson \cite{2017_Thompson_PhysRevLett} Hamiltonians in \fig\ref{fig:figure_9_sup_neutron_field_dependence_vs_theory}.
The diffraction data were acquired at \SI{0.1}{K} (within the ordered phase) and at \SI{5}{K} to show the paramagnetic background. The low temperature data was taken before the high temperature data for cryogenic convenience, so the peaks had to be reacquired via rocking scans. This unfortunately means that some of the data was taken slightly off-peak such that the high temperature data does not precisely match the intensities of the low temperature data---(004) in particular. Nevertheless, all the paramagnetic background scans (green data in \fig\ref{fig:figure_9_sup_neutron_field_dependence_vs_theory}) show a flat field dependence as expected, so the field dependence of the low temperature scattering is due to magnetic changes in the sample itself.

The calculated magnetic scattering is based on the spin structures arrived at via mean-field theory described in the text. To compute scattered intensity, we also included domain selection effects: between 0 and \SI{0.1}{K} (where the internal demagnetizing field is zero), we interpolated between a zero-field state of equal domain population with net ferromagnetic order along $[100]$, $[010]$, and $[001]$, to a \SI{0.1}{K} state including only $[100]$ and $[0 \bar1 0]$. (This is justified by the comparison to the Potts model---see the main text.) This resulted in some fairly dramatic predicted low-field dependence on the $(1 \bar1 3)$ and $(004)$ peaks, shown in \fig\ref{fig:figure_9_sup_neutron_field_dependence_vs_theory}(h) and (l).
From \SI{0.1}{K} and above, we assumed equal population of domains along $[100]$ and $[0 \bar1 0]$.

\begin{figure}[htbp]
	\centering
	\includegraphics[width=\columnwidth]{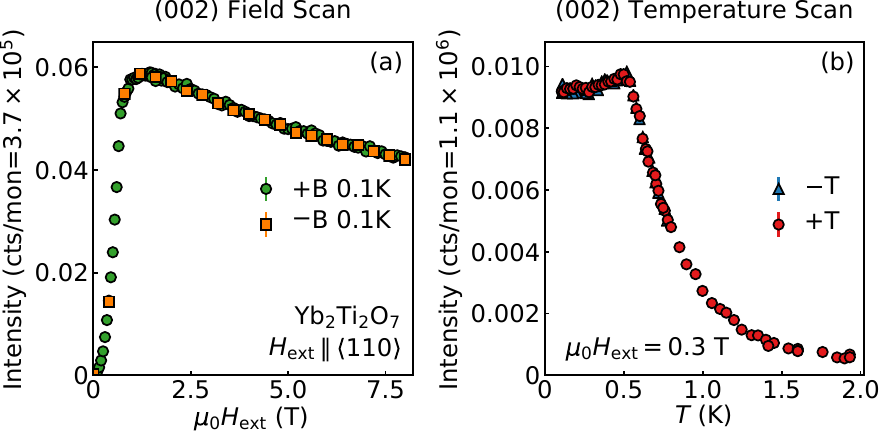}
	\caption{High field and high temperature scattering from the (002) \ce{Yb2Ti2O7} peak. The high field scattering at 100~mK in panel (a) shows that an applied field of \SI{8}{T} does not produce a collinear spin structure. The high temperature scattering in panel (b) shows that an increase in temperature reduces the sublattice magnetization as expected. Error bars represent one standard deviation.}
	\label{fig:figure_10_sup_002_highBhighT}
\end{figure}

Given that the (002) intensity indicates a non-collinear spin structure, it is worth asking how high of a magnetic field would produce a collinear polarized spin structure. Scattering in a $\langle 110 \rangle$ field up to \SI{8}{T} is shown in \fig\ref{fig:figure_10_sup_002_highBhighT}(a), and shows only a modest decrease in intensity from the maximum value around \SI{1}{T}. This indicates that the spin structure remains non-collinear up to applied magnetic fields in excess of \SI{8}{T}, as one would expect given that the lowest energy excited crystal field level is at $\sim 60\,$\si{meV} \cite{2015_Gaudet_PhysRevB}. 
When we increase temperature on the (002) peak, we see a steady decrease in intensity as the thermal fluctuations diminish the sublattice magnetization. This is shown in \fig\ref{fig:figure_10_sup_002_highBhighT}(b).

Qualitatively, all the theoretical calculated intensities are the same. They only differ in relative intensity and the upper critical field. Given that the upper critical field is renormalized by quantum effects \cite{2019_Rau_PhysRevB}, comparisons of critical field are not a good way to adjudicate between the proposed Hamiltonians.
However, for the (111), (222), (113), and (004) peaks, the theoretical calculated intensity does not even resemble the experimental data. In each of these cases, there is a sudden drop in intensity at low magnetic fields which is either not expected or the opposite of what is expected for domain selection. A possible explanation for this is magnetic extinction. Extinction in ferromagnets is reduced when there are many domain walls, but then enhanced when a field reduces the number of domain walls. This causes a sudden drop in scattered intensity when a magnetic field is applied, as seen in yttrium iron garnet \cite{1989_Baruchel_PhysicaBCondensedMatter}. This precisely matches what we observe in \ce{Yb2Ti2O7}. Thus, only the weaker Bragg peaks (002) and (220) have scattering intensity which resembles the data. As explained in the text, these data indicate a non-collinear $\langle 110 \rangle$ field-polarized phase due to easy-axis single ion anisotropy.

\subsection*{Coarse-Grained Model}

The cubic anisotropy is minimized with a six-fold degeneracy for magnetization along $\{ \pm \vec{e}_i \}$, \ie $\vec{e}_i \,=\, \hat{\vec{x}}, \hat{\vec{y}}, \hat{\vec{z}}$ in the global frame. Ignoring further high-order terms, the potential energy for the magnetization represented by a unit vector $\vec{m}$ is 
\begin{equation}
\label{eq-sub:anisotropy-zeeman-potential}
U = -K_1 \sum_i (\vec{m} \cdot \vec{e_i})^4 - K_2 \prod_i (\vec{m} \cdot \vec{e_i})^2 - \vec{h} \cdot \vec{m}.
\end{equation}
In the limit of a strong magnetic field, the spin is fully polarized along the direction of the field $\vec{a}_3$. Near and below the transition field, $\vec{m}$ develops a small deviation $\rho$ in the two transverse directions $\vec{a}_1$ and $\vec{a}_2$. In the local frame, $\vec{m} \,=\, \rho (\vec{a}_1 \cos \phi + \vec{a}_2 \sin \phi) + \vec{a}_3 \sqrt{1-\rho^2}$.

For a magnetic field of magnitude $h$ applied along $\langle$110$\rangle$, the local frame is defined by 
$\vec{a}_1 \,=\, (0,0,1)$, $\vec{a}_2 = (1,-1,0)/\sqrt{2}$, and $\vec{a}_3 = (1,1,0)/\sqrt{2}$. Expanding the potential energy gives, up to addition by a constant,
\begin{align}
U^{\langle 110 \rangle} \,=\, \bigg[ &\left(\frac{h}{2} - \frac{K_1}{2} - \frac{K_2}{8} \right) + \nonumber  \\
									 &\left(\frac{3 K_1}{2} - \frac{K_2}{8}\right) \cos 2 \phi \bigg] \rho^2 + O(\rho^4).
\end{align}
For the range of values of $K_1$ and $K_2$ we are working with, the minimization with respect to $\phi$ gives $\cos 2 \phi \,=\, -1$ and the coefficient for $\rho^4$ is positive definite. At $h_c^{\langle 110 \rangle} = 4 K_1$, the minimum at $\rho = 0$ becomes unstable, giving a second order phase transition. 

Approaching the transition field from below with $ h  \,=\, h_c^{\langle 110 \rangle} - \delta h$,
\begin{equation}
U^{\langle 110 \rangle} \,=\, - \frac{\delta h}{2}\rho^2 + \left(\frac{5K_1}{2} -\frac{\delta h}{8}\right)\rho^4 + O(\rho^6).
\end{equation}
Thus, the magnetization along the field scales with $\delta h$ linearly,  
$m \,=\, \vec{m} \cdot \vec{a}_3 \,\approx\, 1 - \delta h/ (20 K_1) $, until the slope suddenly jumps to 0 for $h \,\ge\, h_c^{\langle 110 \rangle}$.

For the field $h$ along $\langle 111 \rangle$, the local frame is given by $\vec{b}_1 \,=\, (1,-1,0)/\sqrt{2}$, $\vec{b}_2 \,=\,, (1,1,-2)/\sqrt{6}$, and $\vec{b}_3 \,=\, (1,1,1)/\sqrt{3}$. Similarly, up to a constant term, the potential energy is 
\begin{align}
U^{\langle111\rangle} \,=\, &\left(\frac{h}{2} - \frac{4K_1}{3} + \frac{2K_2}{9} \right)\rho^2 \nonumber \\
						-&\left(\frac{2\sqrt{2}K_1}{3} + \frac{\sqrt{2}K_2}{27}\right)\rho^3 \sin 3 \phi \nonumber \\
						+&\left(\frac{h}{8} + \frac{7 K_1}{6} - \frac{5K_2}{12}\right)\rho^4 +  O(\rho^5),
\end{align}
which we denote as 
$U^{\langle111\rangle} \,=\, c_2\rho^2/2 - c_3\rho^3/3 + c_4\rho^4/4 +  O(\rho^5)$.

Approaching the transition field from above, we expect a first order phase transition at $h_c^{\langle111\rangle}$ that satisfies $2 c_3^2 \,=\, 9 c_2c_4$, where $\rho$ suddenly develops a finite value $2c_3/3c_4$, accompanied by $\sin 3\phi \,=\, 1$, giving a divergent slope in the magnetization $m = \vec{m} \cdot \vec{b}_3$.

\end{document}